# Physical and Mechatronic Security, Technologies and Future Trends for Vehicular Environment

## Towards Counteracting Cloning in Automotive Industry

Prof. **Wael Adi,** Dipl.-Ing. **A. Mars,**
IDA, Technische Universität Braunschweig;

## 1. Abstract

Cloning spare parts and entities of mass products is an old and serious unsolved problem for automotive industry. The economic losses in addition to loss of know-how and IP theft as well as security and safety threats are huge in all dimensions. This presentation gives an overview on the traditional state of the art on producing clone-resistant electronic units in the last two decades. A survey is attempting to demonstrate the techniques so far known as Physically Unclonable Functions PUFs showing their advantages and drawbacks.

The necessity for fabricating "mechatronic-security" in vehicular environment is emerging to become a vital requirement for new automotive security regulations (legal regulations) in the near future. Automotive industry is facing a challenge to produce low-cost and highly safe and secure networked automotive systems. The emerging networked smart traffic environment is offering new safety services and creating at the same time new needs and threats in a highly networked world. There is a crying need for automotive security that approaches the level of the robust "biological security" for cars as dominating mobility actors in the modern smart life environment.

Possible emerging technologies allowing embedding practical mechatronic-security modules as low-cost digital alternative are presented. Such digital clone-resistant mechatronic-units (as Electronic Control Units ECUs) may serve as smart security anchors for automotive environment in the near future. First promising initial results are also presented.

## 2. Introduction

The Physically Unclonable Functions (PUFs) were introduced two decades ago for fabricating electronic unclonable units for secured identification or for intellectual property protection [1] to [31]. Due to the analog nature of all proposed PUF technologies, virtually all techniques proposed so far had limited use in real world applications due to economic cost factors and failing long term stability. The majority of such PUF techniques offer idealized DNA-like identity for a physical entity. However, all such technologies suffer from being highly inconsistent due to being sensitive to aging, temperature, power fluctuations and other operational conditions. The objectives of this presentation are to show the different known traditional unclonability technologies and their drawbacks for vehicular industry.

A secured "Mechatronic-Identity" is one of the most wanted requirements in automotive systems for producing unclonable or clone resistant units and spare parts. This technology seems to be in its childhood due to the expected high cost factors when embedded in vehicular units like Electronic Control Units (ECU) and seriously more problematic in mass-products as automotive wearing-parts. In the contemporary emerging world of Internet of Things (IoT), automotive units are becoming even a part of the worldwide communicating networked units. This requires however robust, consistent and low-cost technologies for safe and secure operation. This is admittedly a very challenging research and development area addressing the "Physical Security" issues which started recently to grow up exponentially in its importance in the research environment. Unclonable uniqueness is expected to play a major role as a security anchor in all our future applications; similarly as it is the case in our well established smart biological environment. Technology will possibly not attain the quality of biological DNA-identity. Modern technology can however, try to develop bio-inspired techniques.

The following sections include first an excursion in the world of PUF technologies followed by introduction to some basic concepts of mechatronic-security. Finally a summary of some research activities on *"Mechatronic-Security"* and "practical Digital-PUFs" at IDA, Institut für Datentechnik und Kommunikationsnetze of the Technical Universität Braunschweig are presented. This is an attempt to initiate hopefully fruitful discussions and mutual exchange of ideas and experiences between industrial and academic communities.

### 3. Basic Unclonability Concepts and Physically Unclonable Functions PUFs

Physical Unclonable Function (PUF) is a function embodied in a physical unique structure where its output looks like a random function, it is easy to evaluate but hard to predict even for an attacker with physical access. PUFs are increasingly used in cryptographic systems especially for devices identification/authentication, secure memoryless key storage, IP protection and anti-counterfeiting. As unclonable entities, PUFs should fulfil the following security requirements:

**Evaluable:** For a given PUF, it should be easy to get a response R for any challenge C, R=PUF(C). The C-R pair is called a challenge-response pair CRP.

**Uniqueness:** PUFs should be unique, such that when challenging different PUFs with the same challenge C, each PUF generates a unique response which is also different from all other units. Fig. 1 illustrates this property.

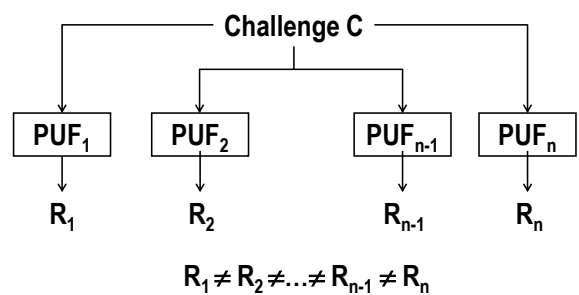

Fig. 1. PUFs uniqueness property

**Robustness:** Means that the PUF should generate consistent and repeatable responses within the whole system lifetime. Deviating responses should be impossible or tolerable by some adequate means.

**Unclonability:** It should be intractable for an adversary (including the manufacturer) even when physically attacked to create a physical or software clone of the PUF.

**Unpredictability:** The adversary should not efficiently be able to compute the response of a PUF to any challenge, even if he/she can adaptively obtain unlimited number of CRPs from the same PUF or other instances.

**One-way:** PUFs are difficult to invert such that given a PUF and one of its responses R, it is hardly infeasible to find the corresponding challenge C.

**Tamper evident:** The PUF should be tamper-evident. That is by any attempt to physically access the PUF, the PUF would changes its challenge/response behavior.

## 4. PUF instantiations

The idea of using the physical unique structure to identify units is similar to fingerprint identification of human beings. In the twentieth century, non-electronic PUFs have been introduced; they were using random patterns on paper and optical entities to extract unique identities. Recently, many electronic technologies for PUFs instantiations have been proposed, Fig. 2 shows a classification table for the most known PUFs instantiations.

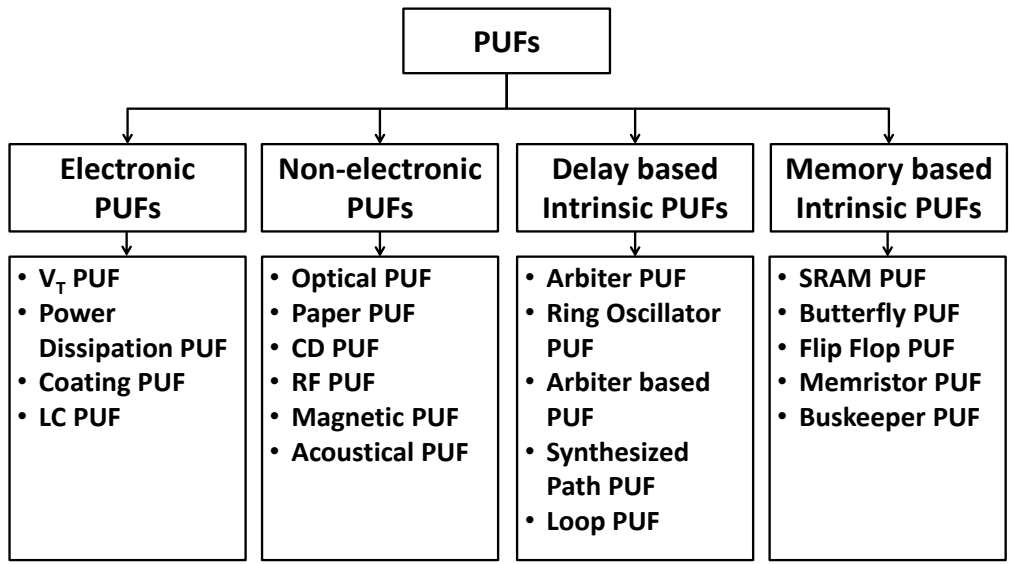

Fig. 2. Physical Unclonable Functions instantiations

### 3.1. Electronic PUFs

#### 3.1.1. $V_T$ PUF

$V_T$ PUF was introduced in [1], it consists of an addressable array of equally designed transistors driving resistive loads. The randomness is coming from the variations of the threshold voltage ($V_T$) of each transistor, the addressed transistor drives a resistive load, and the resulting random analog voltage sequence is converted to a binary identification sequence.

#### 3.1.2. Power Dissipation PUF

Power Dissipation PUF was introduced in [2], it is based on the resistance variation of the power grid of chips which is connected to Power Ports (PP). The response consists of a set of voltage drops at a set of distinct locations. The measured electrical parameters are random and unique. Fig. 3 describes a sample instrumentation setup of Power Dissipation PUF for 6 PPs..

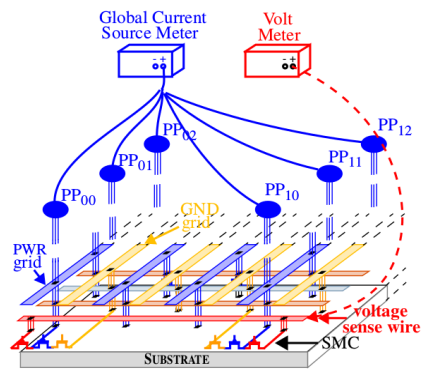

Fig. 3. Instrumentation setup of Power Dissipation PUF [2]

### 3.1.3. Coating PUF

Coating PUF was introduced in [3]. A network of metal wires (sensors) is laid out in a comb shape. The space between and above is filled with opaque material and randomly doped with dielectric particles. The unpredictable capacitance values are measured with comb-shaped sensors on the top metal layer of an integrated circuit.

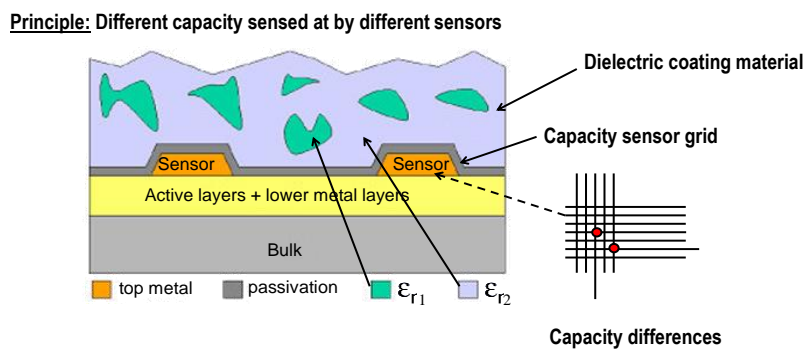

Fig. 4. Dielectric Coating PUF

### 3.1.4. LC PUF

LC PUF was introduced in [4]. A passive resonator circuit (LC circuit) absorbs an amount of power when a RF electromagnetic field is generated. The power depends on the frequency and of the precise characteristics of the capacity and inductance of the LC circuit that uniquely identifies an LC circuit.

### 3.2. Non-Electronic PUFs

### 3.2.1. Optical PUF

Optical PUFs have been proposed first in [5], where reflective particle tags were developed for uniquely identifying strategic weapons. In [6], optical PUFs were proposed as Physical One-Way Functions (POWF). The concept of optical PUF is presented in Fig. 5. A Laser beam is directed to a light scattering material. Random and unique speckle pattern will arise, the pattern is captured by a camera for digital processing to generate a unique identity. To increase the number of CRPs, different laser orientations can be used.

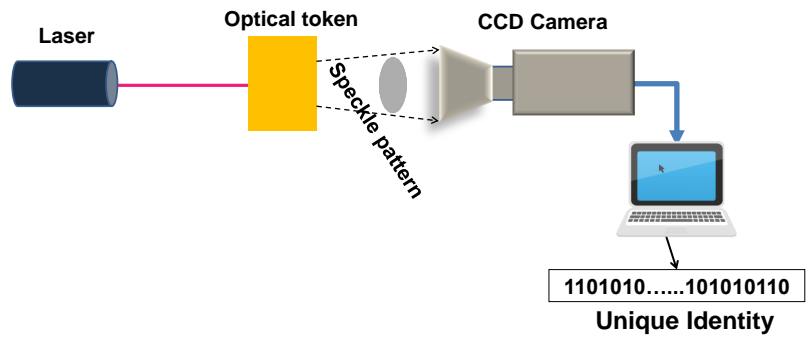

Fig. 5. Concept of Optical PUF

### 3.2.2. Paper PUF

Paper PUF was proposed in [7], It consists of scanning the unique and random fiber structure of regular or modified paper, the reflection of a laser beam on the random fiber structure is used as a fingerprint.

### 3.2.3. CD PUF

The data is stored as a series of lands and pits formed on the surface of the CD. It was observed that the measured lengths of lands and pits of a regular CD contain unpredictable random deviations which can be used as an identification profile [8].

### 3.2.4. RF DNA

In one implementation scenario, a physical token as a mixture of conducting and dielectric materials integrated with an RFID is produced. A high frequency wave (5-6 GHz) is propagated through the token. A low-cost antenna matrix receives a fingerprint of the token and use it as an identification profile [9].

### 3.2.5. Magnetic PUF

Magnetic PUF is based on determining the remanent noise in a magnetic medium by DC (Direct Current) saturation of a region thereof and measurement of the remaining DC magnetization. The remanent noise may then be digitized and recorded on the same magnetic medium to thereby "fingerprint" the magnetic medium [10].

### 3.2.6. Acoustical PUF

Acoustical PUF presented in [11], is based on the random delay in a fiber-glass delay line integrated in a device (DL701) used in televisions. When probing DL701 with one certain caustic wave, the internal medium of DL701 changes the wave response characteristics individually. In [11], it is shown that each individual DL701 produces a unique response. To create large number of CRPs, each unit is probed with different frequencies. Fig. 6 describes the principle of the acoustic delay line PUF concept of DL701.

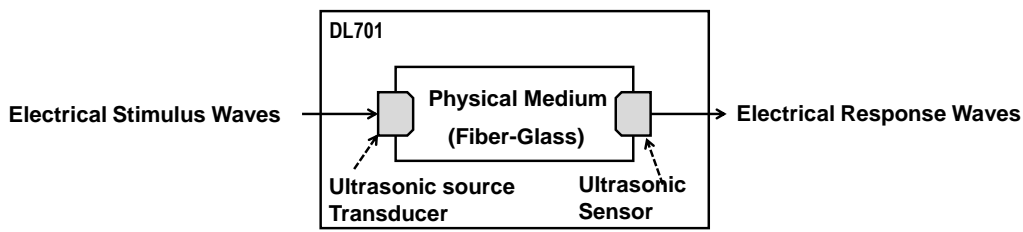

Fig. 6. Principle of DL701 Acoustic PUF

### 3.3. Delay-Based Intrinsic PUFs

#### 3.3.1. Arbiter PUF

The initial proposal of Arbiter PUF was introduced in [12][13], it is based on the random delay or latency of each switch block, such that for two symmetrically designed units, and because of the manufacturing variations, there is a small random offset delay between the delays of the two units. For n switch blocks, there exist $2^n$ different delays selections as a stimulus. Fig. 7 illustrates the principle of an arbiter PUF.

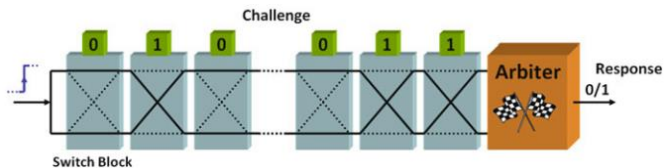

Fig. 7. Basic concept of arbiter PUF [14]

#### 3.3.2. Ring Oscillator RO PUF

RO PUF is also based on random delays caused by manufacturing variations. It is based on a delay line where its output is inverted and fed backed to its input, resulting with an asynchronously oscillating loop. The PUF response is extracted by measuring the frequency of the RO, where a simple edge detector can detects the rising edges in the periodical oscillation and a counter counts the number of edges over a period of time [15]. Ring Oscillator structure is also practically deployed in Microsemi S devices as a true random entropy source for the Number Random Bit Generator (NRBG), it provides 384 full entropy conditioned bits to the Deterministic Random Bit Generator (DRBG) [16].

#### 3.3.3. Arbiter-Based PUF

APUF was introduced in [17] to enhance the reliability of delay-based PUFs against temperature variations. APUF compensates the thermal-induced delay variation in the PUF circuits; a Voltage Controlled Current Starved (VCCS) inverter chain is regulated by adapting its control voltage by a complementary to absolute temperature (CTAT) reference generator.

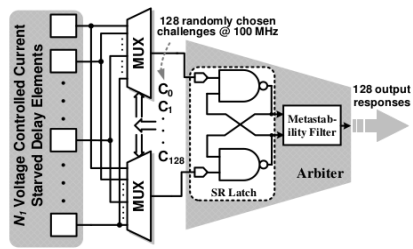

Fig. 8. Block diagram of APUF [17]

#### 3.3.4. Loop PUF

Loop PUF was introduced in [18], it has nearly the same principle as the RO PUF. N identical and controllable delay chains are connected in a loop to create a ring oscillator, each delay chain contains M delay elements as shown in Fig. 9. The controller applies different combinations of the M control bits and measures the frequency or the delay of the oscillating loop.

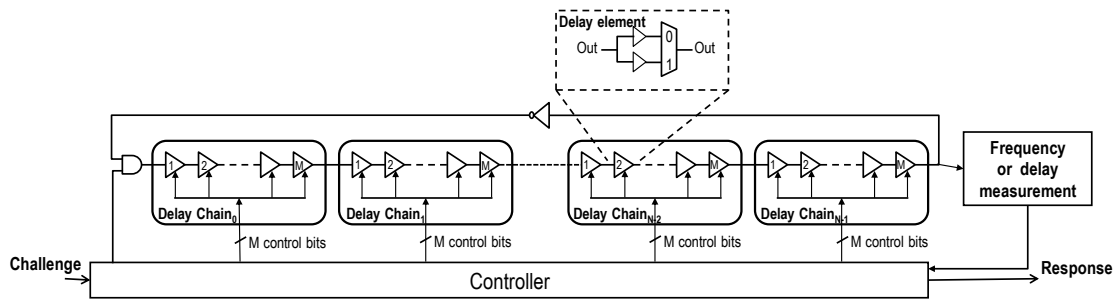

Fig. 9. Loop PUF structure

### 3.3.5. Sensitized Path PUF

SP-PUF was introduced recently in [19], an existing circuit is turned into an SP-PUF. It is measuring the delay differences in selected delay paths in existing designs by adding Race-Resolution Element (RRE) and multiplexers MuxA and MuxB as described in Fig. 10 [19].

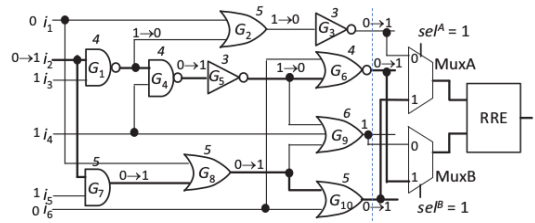

Fig. 10. Example of SP-PUF [19]

### 3.4. Memory based intrinsic PUFs
### 3.4.1. SRAM PUF

SRAM PUFs is one of the most practically used PUFs. was introduced in [20], this research was carried out at Phillips Research, before the company Intrinsic ID was founded. Intrinsic ID [21] created SRAM PUF, based on the behavior of standard SRAM memory in digital chips. It can differentiate devices such as microcontrollers from each other. Every SRAM cell has its own preferred state every time the SRAM is powered on, resulting from random differences in the threshold voltages. This randomness is expressed in the startup values of "uninitialized" SRAM memory. Hence an SRAM response yields a unique and random pattern of 0's and 1's. This pattern can represent a chip's fingerprint, since it is unique for a particular SRAM and hence for a particular chip. In [14], security analysis of the most popular intrinsic electronic PUFs have been done, resulting that SRAM PUFs have a Bit Error Rates (BER) of about 6% at +25 °C, 8% at -40°C and at +85°C. Also, the BER at 1.20V is about 6% also for 1.32V. SRAM PUFs are used in some Microsemi FPGAs such as SmartFusion®2 SoC FPGAs [16] and IGLOO2 [22] FPGAs that can be used for providing secure memoryless keys for cryptographic applications (PUF unit complexity is around 100 K-Gate Equivalent!). SRAM PUF needs its own protected design area/location and can't use the internal FPGA SRAM resources, as usable SRAMs are hard-resetted to all-zero after power up and hence its randomness is lost. Furthermore, SRAM PUF needs controllable power-up-event to enable the response generation which is not acceptable for all applications.

### 3.4.2. Butterfly PUF

Since SRAM PUFs can't be used in FPGA environment without additional external resources, butterfly PUF was introduced in [23] to overcome the SRAM PUF drawbacks. It imitates the SRAM PUF in FPGA environment without the need for actual device power up, it consists of two cross-coupling latches. By using clear/preset functionality, a random state will be generated depending on the physical mismatch between the latches and the cross-coupling interconnect.

### 3.4.3. Latch PUF

Latch PUF [24] is very similar to Butterfly PUF, two NOR gates are cross-coupled to a simple NOR latch. Depending on the internal mismatch, it converges to a stable state.

### 3.4.4. Flip-Flop PUF

Flip-Flop PUF was proposed in [25], it is based on power up characteristics of uninitialized flip-flops. Flip-flops have the advantage of being able to easily spread over an IC which makes them difficult to locate by an attacker.

### 3.4.5. Memristor PUF

Memristor PUF was introduced in [26], memristors are emerging as next generation non-volatile memory technologies, a memristor is defined at logic 0 when $0 < w/D < O_L$ and for logic 1 when $O_H < w/D < 1$, the region $0 < w/D < O_L$ is undefined. Fig. 11 presents the memristor device model.

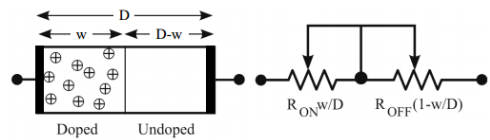

Fig. 11. Memristor device model [26]

The memristor PUF mechanism exploits the unpredictable state of the memristor within the undefined region, where the state depends on the duration of access time to the memristor and the value of supply voltage. Each physical memristor unit behaves individually differently to the same challenge that can be a Short Write Time (SWT) or Low Write Voltage (LWV).

### 3.4.6. Buskeeper PUF

Buskeeper PUF was proposed in [27], Buskeeper or Busholder is a week latch that usually has no control signals. It is intended to be used with on-chip buses that have multiple drivers, it is equivalent to a DFF with the enable signal connected to Vdd [27], also it has lower hardware complexity than Latches and DFFs. The principle of Buskeeper PUF is similar to all memory based intrinsic PUFs where the initial patterns are read at the memory start up.

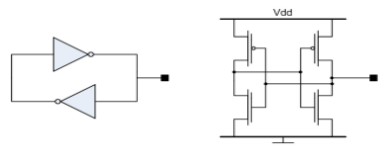

Fig. 12. Buskeeper cell structure [27]

### 5. Fuzzy extractors

PUFs are inconsistent in their behavior because of their sensibility to the environmental conditions variations such as temperature, voltage variations, radiation deviations and aging factors. To overcome the PUFs inconsistency, Fuzzy Extractors or helper data algorithms are used to extract consistent responses from the PUFs. Fuzzy Extractors are designed to correct 25% of inconsistency errors or more. Fuzzy extractors induce however new security weakness that will be discussed in the following section. Fig. 13 describes the general constellation strategy of PUFs with fuzzy extractors.

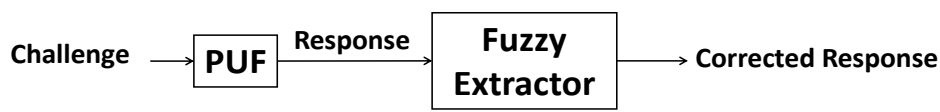

Fig. 13. Use of Fuzzy Extractor with PUFs

## 6. Cloning Attacks on Physically Unclonable Functions

Two types of cloning are well known: first; Mathematical/modelling cloning that aims to create an algorithm that behaves similarly as the targeted PUF. Second; A Physical cloning characterizes the physical response of the targeted device and creates an identical physical response in a second instance of the same device type.

The process of cloning a PUF consists of two steps [28]:

- **Characterization:** a process in which the attacker gains knowledge of the challenge/response behavior of a PUF.
- **Emulation:** the process of recreating or modelling the unique response of a PUF, i.e. creating a PUF with identical challenge/response pairs.

In [29][30], authors present successful modelling attacks on several known strong PUFs, including Arbiter PUFs, XOR arbiter PUFs, Feed-Forward Arbiter PUFs, Lightweight Secure PUFs, and Ring Oscillator PUFs.

In [31], side-channel leakage attacks on PUFs and Fuzzy Extractors were analyzed. Attacks targeting weak PUFs and their fuzzy extractors were introduced. The presented analysis covers Arbiter PUFs and RO PUFs, by deploying side-channels mainly power consumption and electromagnetic emission. For fuzzy extractors, Simple Power Analysis (SPA) was used to attack both Code-Offset fuzzy extractor and Toeplitz hashing; hence extracting the cryptographic key derived from PUFs structure was possible.

In [28], cloning SRAM PUF by side channel analysis was treated. As SRAM PUFs use standard on-chip memory interfaces and buses, attacker can gain control of such interfaces, where he/she could read the memory content and hence clones the SRAM PUF. Invasive de-capsulation with micro-probing can provide access to any memory content, mostly if the memory IC is separated. For SRAM PUFs embedded in FPGA for example, this method is infeasible because of the huge number of interconnections. Several Side Channel Analysis (SCA) techniques can be used to extract the memory content or part of it, if the targeted device includes an inspection resistant memory. In [28], Photonic Emission Analysis (PEF) was used to extract the full content of an SRAM embedded in Atmel ATmega328P. PEF is passive, non-destructive and a semi-invasive SCA. The proposed attack uses a backside approach to clone the SRAM startup behavior exploiting the most known countermeasures seeking to detect malicious modifications from the chips front sides. The amount of lab time necessary to produce an initial clone was about twenty hours, and subsequent clones was produced more easily in less than three hours [28].

## 7. Mechatronic Security

### 7.1. Optimized Secured Mechatronic Identity Model

Linking electronic units to mechanical units such as sensors or actuators, results with joint entities assigned as mechatronic units. The security of such entities starts by securely identifying such units as unique entities. Therefore, a secured mechatronic identity is a first step towards cloning protection in automotive environment. Fig. 14 shows a possible joint structural mechatronic identity. One ultimate implementation is by embedding the electronic unit in the physical structure when possible. The system includes now two identities joined together, namely the electronic one and the identity extracted from the body of the physical structure (physical structure-DNA itself). In that case the resulting unit becomes unclonable as any invasive physical attack on the electronic unit would destruct the identity profile and hence destructs the secret to be cloned in the physical body.

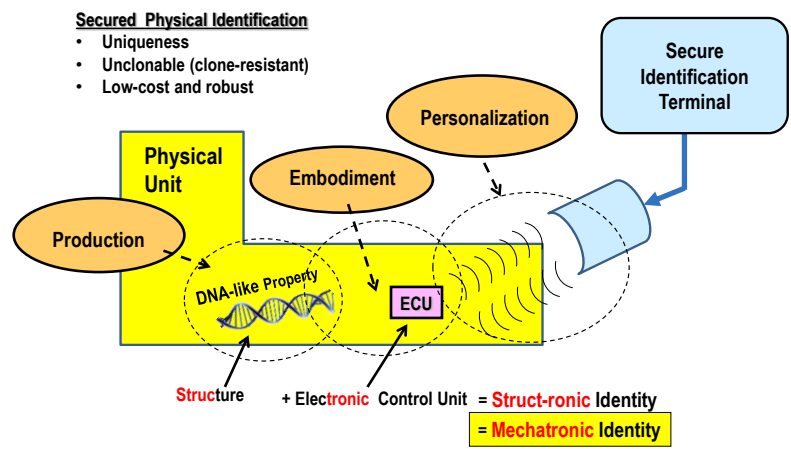

Fig. 14. Conceptual Mechatronic Security

### 7.2. Sample Implementation Scenario for a Mechatronic Identity

One possible DNA like identity profile is the amorphous material structure of a physical body. Such structures exhibit properties which are impossible to predict, copy or to model. These can be adequate structural properties to be used as a part of a DNA-like identity chain.

To extract randomly a part of the internal structure, many mapping stimulation can help to get responses to some randomly selected stimulus from a huge stimulus class. If the number of stimulus possibilities exceeds say $2^{80}$, then the trace of that structure is deemed as impossible to clone.

Possible stimulus technologies on physical structure can be such as (but not limited to): Optical, acoustic, electrical, chemical, etc. which can be used jointly or individually as a Structure-DNA.

Fig. 15 shows a possible sample conceptual DNA-like profile extraction constellation by using ultrasonic stimulation wave. The wireless link to the electronic control unit ECU is selected just to offer optimized security level against invasive attacks. The reason is that, any attempt to mechanically reach the ECU would destruct the secret structural response and hence destruct the DNA-like response and finally destruct the identity. That is, cloning becomes impossible by any invasive attack!

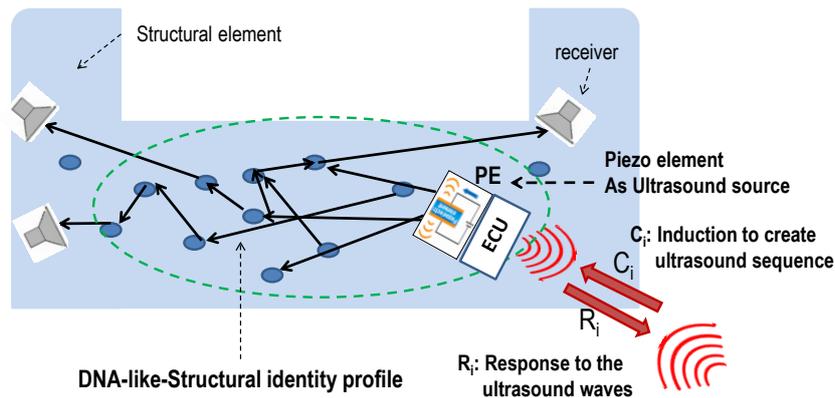

Fig. 15. Conceptual Identity Profile Extraction by Acoustic Waves

Fig. 16 shows a possible stimulation train scenario for the constellation of Fig. 15. The first attained challenge –response pair show very promising results for a DNA-like chain with very good cryptographic entropy exceeding 200 bits.

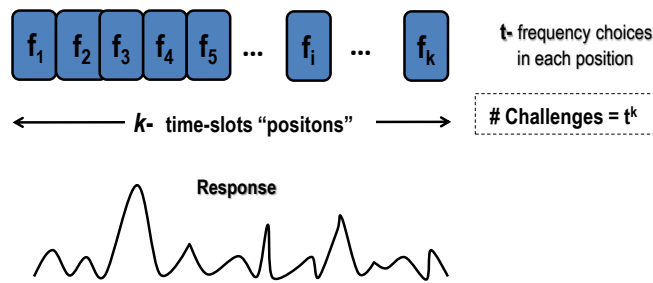

**Cryptographic Entropy Example** :
- For t=32 frequencies in k=20 slots results with an identity space of **Z=$2^{100}$**
- by occupying only p=10 out of k slots results with an identity space of **Z=$2^{65}$**

Fig. 16. Sample Acoustic Wave Stimulation Sequence

When taking into consideration the cancellation of the aging, temperature and other operation condition factors, the entropy may get down tremendously. However, even a low entropy of say 40 bits would still offer acceptable mechatronic security. The reason is that the total identification entropy is the sum of both the structural and the ECU identification entropies. The digital ECU entropy is however highly scalable as would be seen in Section 6.

Fig. 17 shows a first experimental acoustic wave based identification response in two equally fabricated "Eyton" blocks stimulated at a speed of about 40 KHz ultrasonic stimulation impulse.

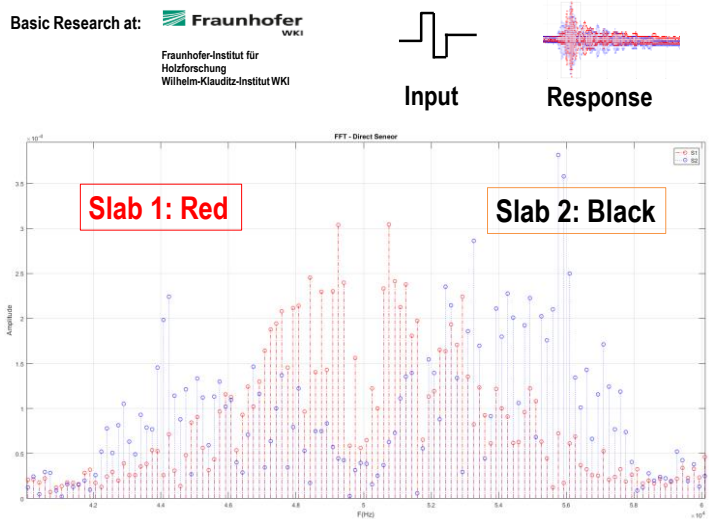

Fig. 17. Experimental Identification Response Differences after DFT Mapping

## 8. Proposed ECU Embedded Digital PUF Concept:

### 8.1. Secret Unknown Cipher

A new digital PUF concept was developed at IDA, Technische Universität Braunschwig since 2008. The main idea behind the concept is to replace the traditional analog PUF technologies mentioned in section 4 by digital PUFs.

The key technique of the proposed digital-PUFs is to create own digital PUF within existing ECUs. The result is a usable and practically implementable security however theoretically not perfect as the analog PUF.

The proposed digital PUF technique is only possible if a self-reconfiguring nonvolatile hard and software technology as an ECU is used. Such technology is expected to be available in the near future.

Fig. 18 shows the digital PUF creation or *"Mutation"* concept as a Secret Unknown Cipher SUC. Notice that the security of the system is not equivalent to "security by obscurity"!!. The reason is that the created/mutated cipher is a random-unknown-cipher and is <u>not known to anybody</u>. The only secret which can be kept absolutely secure is the one which nobody knows.

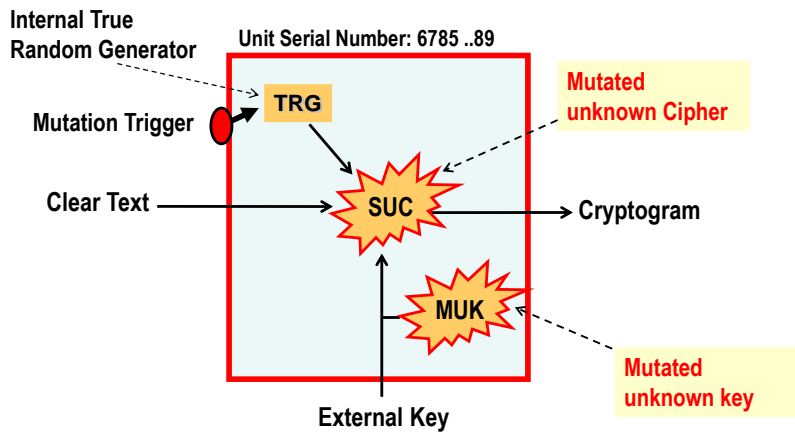

Fig. 18. The Secret Unknown Cipher (SUC) Principle

Fig. 19 shows a practical scenario for creating a SUC by a non-predictable and non-repeatable single-event personalization process in a post-fabrication operation.

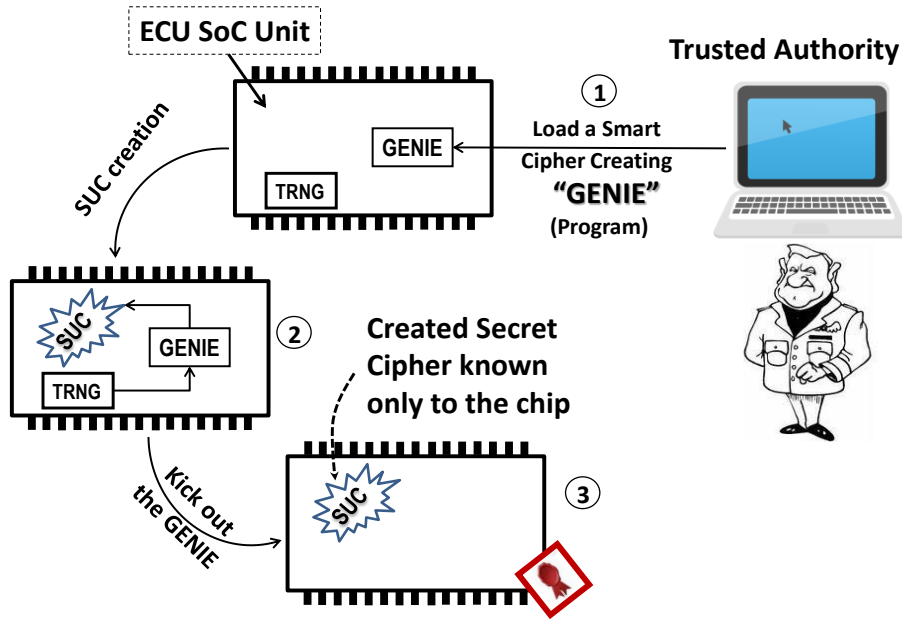

Fig. 19. Embedding Secret Unknown Cipher (SUC) Principle as a Digital PUF in ECU

Fig. 20 shows the personalization and identification protocol of the created individual SUC in a certain ECU.

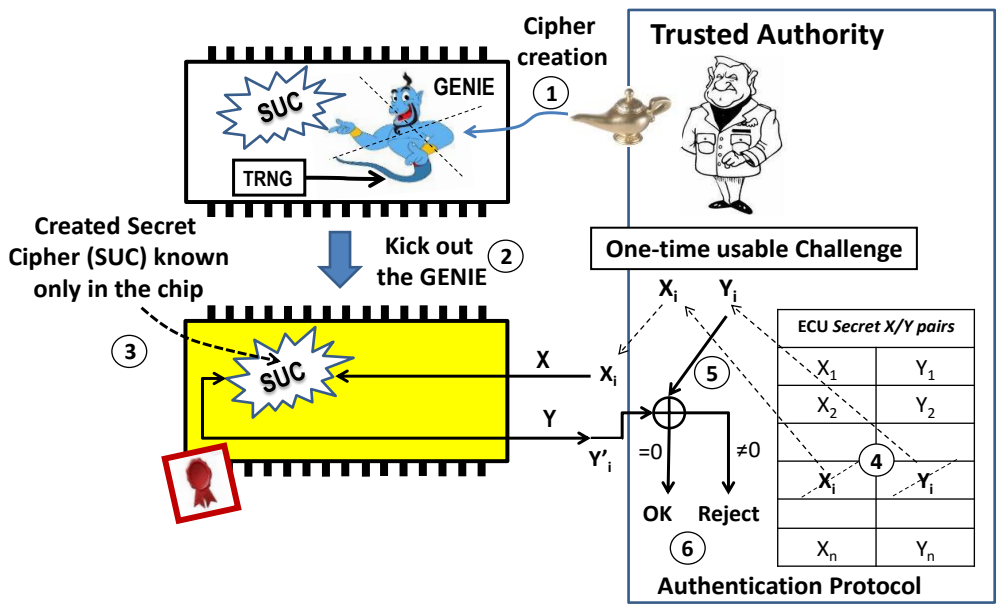

Fig. 20. Key Idea of SUCs in an ECU and a Primitive C-R identification Protocol

**8.2. Sample Digital PUF Prototype as a SUC**

Fig. 21 shows one sample layout implementation of a SUC having an entropy of 80 bits in a non-volatile Microsemi FPGA SoC technology.

The layout shows a very low complexity. In other words, if an ECU using such SoC FPGA (Cost ca. 1 to 3 $), then embedding a secured digital identity in the ECU unit may cost nothing other than a personalization process running for few milliseconds.

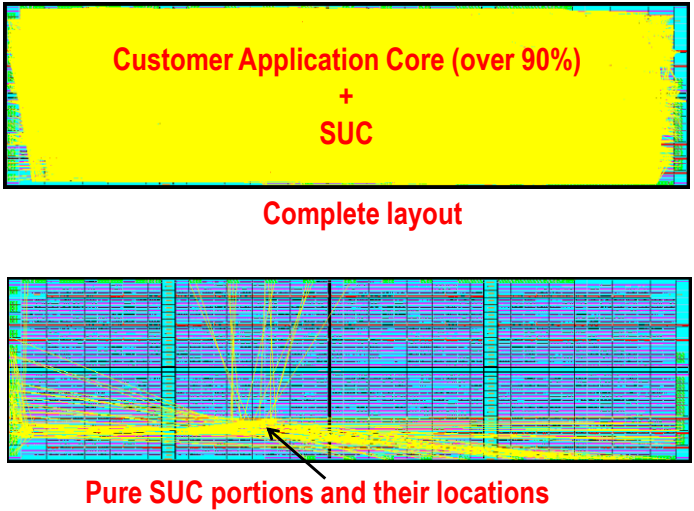

Fig. 21. Sample SUC implementation in a Microsemi SoC FPGA

The accommodated SUC identity has a hardware complexity of just 213 LUTs and 72 register bits. Fig. 21 shows different chip selection scenarios consuming at most 4,5% of the programmable hardware resources of the chip area.

The following figure presents the hardware complexity of SUC (in percent) for different SmartFusion®2 SoC FPGAs Families

**Gate Complexity: 213 LUTs + 72 DFFs**

| SmartFusion2 SoC FPGA Families | Resources usage | |
|---|---|---|
| | LUTs % of usage | DFFs % of usage |
| M2S005 | 3,51 | 1,19 |
| M2S010 | 1,76 | 0,6 |
| M2S025 | 0,77 | 0,26 |
| M2S050 | 0,37 | 0,12 |
| M2S060 | 0,37 | 0,12 |
| M2S090 | 0,24 | 0,08 |
| M2S150 | 0,14 | 0,04 |

**Cipher Cardinality:**
- The cardinality of this class of random cipher is $2^{274}$

**Attack complexity:**
- Linear cryptanalysis: $\geq 2^{80}$
- Differential cryptanalysis: $\geq 2^{80}$

Fig. 22. SUC implementation Complexity in a Microsemi SoC FPGA

Fig. 23 shows a possible practical realization scenario for low-cost clone-resistant automotive units.

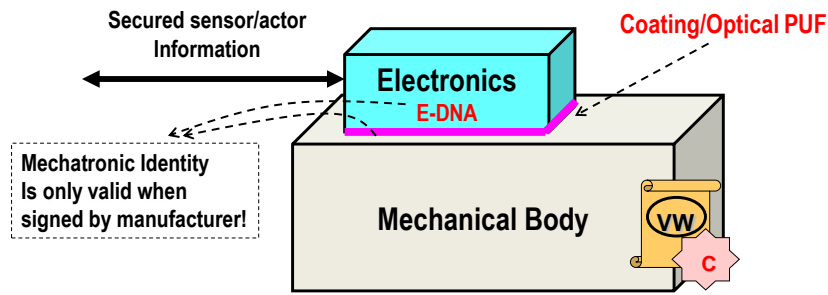

Fig. 23. Low-Cost Secured Automotive Mechatronic Unit

## 9. Conclusion

The emerging vehicular security requirements would require in the foreseen future that each vehicular unit is expected to fulfill the following security requirements:

- Each Car should accommodate unique, unclonable and remotely provable secured Identity
- The car manufacturer should not be able to create the same identity (that is, equivalent to the biological identity)
- The identity technology should be robust and stable for at least 25 years
- Some ECUs should be clone-resistant or unclonable

The required technology is still highly challenging in the contemporary technologies. Mechatronic security (Physical Security) is going to play a fundamental role as a security anchor in all future vehicular systems as DNA in the very robust and reliable natural biological systems.